\documentclass[12pt]{article}
\usepackage{amsfonts,graphics}
\newcommand{\be}{\begin{equation}}
\newcommand{\ee}{\end{equation}}
\newcommand{\bea}{\begin{eqnarray}}
\newcommand{\eea}{\end{eqnarray}}

\newcommand{\ra}{\rightarrow}

\newcommand{\Lra}{\Longrightarrow}
\newcommand{\half}{\frac{1}{2}}
\newcommand{\Half}{\frac{3}{2}}

\begin{document}
 
\pagestyle{empty}
\setcounter{page}{0}

\vspace{20mm}

\begin{center}
{\Large \textbf{A Mathematical Model Accounting for the Organisation  in 
Multiplets of the Genetic Code}}

\vspace{20mm}
{\large  A. Sciarrino }

\vspace{10mm}

\emph{  Dipartimento di Scienze Fisiche,
Universit\`a di Napoli ``Federico II''}

\emph{and I.N.F.N., Sezione di Napoli}

\emph{Complesso di Monte S. Angelo, Via Cintia, I-80126 Napoli, 
Italy} 

\end{center}

\vspace{12mm}

\begin{abstract}
Requiring stability of genetic code   against  translation errors
 , modelised by suitable mathematical operators in the 
 crystal basis model  of the genetic code, the main features of the 
  organisation in multiplets of the
mitochondrial and of the standard genetic code  are   explained.
\end{abstract}

\vfill
\vfill

 \vfill
\vfill

\rightline{e-mail: sciarrino@na.infn.it} 
\rightline{DSF-TH-38/00}
\rightline{math-phys/xxx}
\rightline{ February 2001}

\newpage
\phantom{blankpage}
\newpage
\pagestyle{plain}
\setcounter{page}{1}
\pagestyle{plain}
\setcounter{page}{1}
\baselineskip=17pt


\section{Introduction}

 As it is well known, the DNA macromolecule is 
constituted by two linear chains of nucleotides in a double helix 
shape.  There are four different nucleotides, characterized by their bases: 
adenine  (A) and guanine (G) deriving from purine, and cytosine (C) and 
thymine (T) coming from pyrimidine, T being replaced by uracile (U) in RNA. The 
genetic  information is transmitted  via the messenger ribonucleic acid or 
mRNA. During this operation, called transcription, the A, G, C, T bases in the DNA 
are associated respectively to the U, C, G, A bases. Through a  
complicated biochemical process, a triple of nucleotides or 
codon will be related to an amino-acid.  More precisely, a codon is 
defined  as an ordered sequence of three nucleotides, therefore there are 
$4^{3} = 64$ different codons. Only 20 different
amino-acids appear in the peptide chains which form the proteins. We 
list them  with the standard abbreviation: Alanine (Ala), Arginine (Arg),
 Asparagine (Asn), Aspartic acid (Asp), Cysteine (Cys), Glutamine 
(Gln),Glutamic acid (Glu), 
Glycine (Gly), Histidine (His), Isoleucine (Ile), Leucine (Leu), 
Lysine (Lys), Methionine (Met), Phenylalanine (Phe), Proline (Pro), Serine 
(Ser), Threonine (Thr), Tryptophane (Trp), Tyrosine (Tyr), Valine (Val).
 It follows that the genetic code, i.e. the association between 
codons and amino-acids, is degenerated. In the vertebrate 
 mitochondrial code (VMC) (see Table \ref{tablerep}) 60 of such 
triples  are connected  to 20 the amino-acids,  the remaining 4 codons, 
called non-sense or stop-codons and denoted by the symbol Ter, 
playing the role  to stop the biosynthesis.   
Since the discovery of the genetic code (\footnote{ The literature on 
the genetic code is extremely large. For a recent review with a wide 
selection of references to the original papers  see \cite{Davies}}) a couple of
very puzzling questions have arisen: why only twenty amino acids 
(a.a.) are used in nature to build
up proteins ? why the genetic code  has a peculiar structure in 
multiplets  ranging from  sextets to singlets, in particular: for VMC 
2 sextets, 7 quartets and 12 doublets; for the eukaryotic or standard
genetic code 3 sextets, 5 quartets, 2 triplets, 9 doublets and 2 
singlets ?   An 
attempt to explain the existence of only 20 amino acids is given by the 
hypothesis that
originally the quantum of coding information was transmitted by a pair
of nucleotides instead that by the present triple of nucleotides 
(codon) and $ 4 x 4 = 16$ is a number close to 20. However this 
explanation is in contradiction with other hypothesis on the structure
of the primordial code (quadruples of nucleotides or a subset of the 
present 64 triples). Other explanations are based on correspondence between the 
properties of the amino acids and the structure of the corresponding
codons. Although it seems now clear that a correspondence of this 
kind  exists and it can explain why some amino acids are encoded by more 
codons than others, it is not  evident how the interplay between a.a. and codons leads 
to the existing multiplets structure. A strong and probably correct argument 
  makes appeal to stability considerations, i.e. to state that the
the genetic code has remained unchanged over a vast time period
because it has adopted the most appropriate organisation  to oppose  
the most frequent   and lethal changes.
However no consistent model, at my knowledge, has been proposed  
to  explain the actual multiplets structure in the light of
the above statement. It is known that the translation errors are the 
main source of devastating effect in the construction of the polypetide 
chains. Errors in reading the nucleotide in 3rd position are more frequent 
than errors in reading the nucleotide in 1st position and the latter 
are more frequent than those in 2nd or central position.
  Clearly  a protection against the translation errors represented
   by transitions, i.e. the replacement by a pyrimidine (purine) by 
the other pyrimidine (purine), which are  the most common mutations,  is
obtained by encoding the same a.a. by codons of the form XZY or XZR. In the following   
the standard  notation is used:
  \be
  X,Z,N = C,U,G,A \;\;\;\;\;\;\; Y = C,U  \;\;\;\;\;\;\; R = G,A
  \ee   
 Codons encoding the same a.a., i.e. belonging to the same multiplet
 are called  synonymous. Similarly a.a. encoded by codons of the form XZN are, 
 in some way, protected by the  effects of the translation errors  represented by 
tranversions (pyrimidine into purine 
 or  viceversa).  But why there are only six or five quartets ? why 
do only two or three  sextets appear ? why the quartets and sextets have the 
structure they have for the first two nucleotides XZ ? 
 The complete stability against  reading errors is in obvios conflict 
with the advantage to encode  many  a.a., so to allow a very large variety of 
byosynthesis products. 
 It is the aim of this paper to propose a mathematical model which
 may explain both the number of the natural amino acids and the 
 structure in multiplets. The framework in which the model is proposed
 is the crystal basis model of the genetic code \cite{FSS1} in which
 the 4 nucleotides  are assigned to the 4-dim irreducible fundamental 
 representation (irreps.) $(1/2, 1/2)$ of $U_{q \to 0}(sl(2) \oplus 
sl(2))$ with the following assignment for the values of  the third 
component of  $\vec{J}$ for the two $sl(2)$ which in the following will be denoted
   as  $sl_{H}(2) $ and $sl_{V}(2) $ :
\be 
	\mbox{C} \equiv (+\half,+\half) \qquad \mbox{T/U} \equiv 
(-\half,+\half) 
	\qquad \mbox{G} \equiv (+\half,-\half) \qquad \mbox{A} \equiv 
	(-\half,-\half) \label{eq:gc1}
	\ee
	and the codons, triple of nucleotides, to the $3$-fold tensor product of 
 $(1/2, 1/2)$.  We  report   in Table \ref{tablerep} the assignment of the 
 codons to the different irreps.  and the correspondence with the encoded 
 a.a.  in the vertebral mitochondrial  (VMC) and in the standard  
 universal genetic code (SUC). Let us emphasize that the assignments
 of the codons to the different irreps. is a straightforward 
 consequence of the assumed behaviour of the nucleotides eq.(\ref{eq:gc1})
 and of the theorem on the tensor product of irreps. in the crystal basis 
 \cite{Kashi}. The idea of this work is to mathematically represent the effects of 
 translation errors by suitable crystal tensor operator \cite{MS}. Imposing 
 stability of  the genetic code with respect to these errors, i.e. that codons which are 
 most  sensitive to be read in a wrong way correspond to synonymous codons in the 
  encoding process, we find the main features of the multiplet structure of the VMC, which 
  is believed to represent a primordial form of the code, and of SUC.
  The paper is organised as it follows: in Sec. 2 the general ideas of 
  modelisation of misreading of codons are introduced; in Sec. 3 a detailed 
  discussion of the consequences of the mathematical modelisation is given.
  Indeed in order to study the dependence of the results from the 
   assumptions of the operators mimicking the translational errors, two mathematical 
   schemes are analysed,  discussing which results are model dependent. 
  In Sec. 4 a critical discussion of the obtained results as well as
  some directions for further developments are presented.

 \section{Modelisation of misreading of codons}

 We assume, on phenomenogical grounds,  that there is a hierarchy in 
the  occurrence of translation errors  and, in order of decreasing 
intensity, we consider:  
\begin{enumerate}
 \item  the transitions, in particular $C \ra U$ or $G \ra A$, concerning 
 nucleotides in the 3rd position 
  \item  the tranversions, in particular  $C \ra G$, $U \ra A$ and $C \ra A$, in the 
  nucleotides in 3rd position. 
  \item  the transitions (resp. tranversions) concerning nucleotides 
in 1st position
  \item  the transitions (resp. tranversions) concerning nucleotides 
in 2nd position
  \item  the  mutation induced by the transitions (resp. tranversions) 
   on the first two nucleotides   
  \end{enumerate}
 Transitions (tranversions) of the nucleotide in
 the middle position will be considered far weaker than transitions 
 (tranversions) in other positions  both on
 phenomenogical grounds and on the argument that, in the spirit of the
 hierarchical structure of the intensity of mutations, the change of 
 the other nucleotides  is preferred. 
 Indeed there are phenomenogical arguments, confirmed also by our model, 
 that these changes can be neglected. However we prefer to  discuss
 the translational errors in the above order as it allows the most
 natural introduction of the mathematical structure of the operators
 modelising the errors.
 The hierarchy in the  translation errors mechanisms means that
 a multiplet formed in a level is frozen; in the subsequent levels,
 the  merging of two  whole multiplets in a larger 
 structure is possible, if it is induced by the relevant tensor 
operator. If the transition is allowed only for some
 member of a multiplet, there is conflict between the choice of
 merging the multiplet in a larger one, so decreasing the variety
 of encoded a.a. but increasing the protection or preserving the
 multiplets decreasing the level of protection. In this case,  
 the formation of larger structures will generally take place or not 
according to the rule to protect the weakest codons, i.e. the codons more 
 inclined to be misread.  We assume that misreading of nucleotide C 
or A is the most common. However in the following we shall discuss in 
some detail  each of these case. 
Let us emphasize that we want to build the most simple model in which 
the codons, which are most subject to reading errors,  are synonymous; 
in this spirit the  explicitly analysed transitions ($C \ra U$, $G \ra A$) 
or transversions ($C \ra G$, $U \ra A$, $C \ra A$) have not to be 
considered as the  only possible changes, but as the representatives 
which allow the most simple modelisation. In other words transitions and
  transversions in the reversed directions happen, but the protection
against their effects is assured once the concerned codons belong to the same multiplet.
 We consider only the transversions decreasing or leaving unchanged 
the value of $J_{H,3}$. The tranversion $U \ra G$ implies the increasing 
of one unity of $J_{H,3}$, therefore it is not explicitly considered. 
This is an essentially irrelevant simplification, because it is possible to 
show that a suitable modelisation of this transversion leave the   
obtained results unmodified.
 
 In the following we recall briefly the main properties of the 
    ($q \to 0$)-tensor operators or  crystal tensor operators 
 , \cite{MS}, for a generic $U_{q \to 0}(sl(2)) $. They transform  as
\be 
	J_{3}(\tau_{m}^{j}) \equiv m \tau_{m}^{j} \quad J_{\pm} \, 
	(\tau_{m}^{j}) \equiv \tau_{m \pm 1}^{j}
\ee
Clearly, if $|m| > j$ then $\tau_{m}^{j}$  has to be considered 
vanishing.
The state $\psi_{j_{1} m_{1}}$ will be connected by the ($q \to 
0$)-tensor operator $\tau_{m}^{j}$ to the state $\psi_{JM}$  by the
($q \to 0$)-Wigner-Eckart theorem if
\be
\psi_{JM} = \psi_{j_{1} m_{1}} \otimes \psi_{jm}
\ee
A peculiar feature of the Wigner-Eckart theorem, in the limit $q \to 
0$, is that the selection rules do  depend  not only on the rank of the 
tensor operator and on the initial state, but in a crucial way from the 
specific component of the tensor in consideration. The states $\psi_{JM}$  
can be explicitly computed, up to irrelevant numerical factors, by 
performing the tensor product,
 according to the rules given in \cite{Kashi}, of irreps. $ j_{1}$ 
and $ j $. The tensor product of two irreducible representations in the 
crystal basis is  not commutative, therefore one has to specify which is the 
first representation in  the  product. A final important remark: it is 
clear from the  Table \ref{tablerep} that there are generally more than one 
irrep. labelled by the same value of $(J_{H}, J_{V})$ whose content in the 
constituent nucleotides is different. The transformation properties of a crystal 
tensor operator determine which state  is related to an initial one, only 
according to the irrep. to which the initial state belongs, therefore the 
mathematical modelisation by means of an unique tensor operator is expected to 
be too simple and inadequate. Indeed  the nucleotides are  molecules with very 
different physical-chemical properties, while in the crystal basis model 
they all are handled on the same basis as  vectors of an irreducible 
module. Moreover    it can be expected that some 
reading errors of a nucleotide   depend also from the nature of the 
neighbouring nucleotides. In the following we shall take in some way 
into account this fact by a suitable choice of the nature of the tensor 
operator. Notwithstanding these simplifications of the 
mathematical  modelisation, it is quite amazing how many 
 features of the organisation of the genetic code can be obtained.
  Note that, in order not to overload the notation, we do not 
explicitly specify the action of the operators on the nucleotides, but only 
 their transformation properties under $U_{q \to 0}(sl_{H}(2) \oplus 
sl_{V}(2))$. Hopefully it will be clear from the context which kind of process
 we are considering. 
  
  \section{ Mathematical schemes}
  
  We model the transitions and the  transversions  by the following 
crystal tensor operator, the value of the component being determined by
the labels of the nucleotides, see eq.(\ref{eq:gc1}):
 \be
 C \ra U  \;\;\;\;\; \mbox{or} \;\;\;\;\;  G \ra A   
\;\;\;\;\;\;\;\;\;\; 
\tau_{H,-1}^{1}  \otimes \tau^{a}_{V,0}
 \label{eq:t1}
 \ee
 \be
 C \ra G \;\;\;\;\; \mbox{or} \;\;\;\;\; U \ra A   
\;\;\;\;\;\;\;\;\;\; 
\tau_{H,0}^{b}  \otimes \tau_{V,-1}^{1}
 \label{eq:tr1}
 \ee 
\be
 C \ra A     \;\;\;\;\;\;\;\;\;\;  \tau_{H,-1}^{c} \, \otimes \, 
\tau_{V,-1}^{d}
 \label{eq:tr2}
 \ee 
where the values of $a$, $b$, $c$  and $d$ depend on the position 
inside the codons  of the  misread nucleotide and on the irreps. to which the 
codons belong, see below. 
The above choice  for the horizontal (resp. vertical) part of the 
crystal vector operator in eq.(\ref{eq:t1}) (reps. eq.(\ref{eq:tr1})) is 
indeed the most simple choice according to the change in the labels of the states of  
codons for transitions (resp. transversions). The choice of the rank of the 
vertical (resp. horizontal) part of the crystal operator in eq.(\ref{eq:t1})  (resp. 
eq.(\ref{eq:tr1})), as well as the tensor operator modelising
the tranvsersion C $ \ra $ A,  is someway arbitrary. It is indeed a 
way of taking into account, in  mathematical language, the chemical 
difference between the nucleotides and the difference in the mechanism 
responsible for misreading nucleotides in different positions inside the codons. 
The value of the rank of the operator modelising the translational errors 
in 2nd position will be  generally assumed  larger than the one describing 
errors in 1st position and the latter one will be generally assumed larger than the one 
describing errors in 3rd position, so to model the less frequent 
misreading. In particular, in the scheme we shall discuus more in detail, for 
the transitions the rank $a$ of the "vertical"
tensor operator $\tau_{V,0}^{a}$ will be assumed to be $0, 1, 2$
respectively for transitions in 3rd, 1st and 2nd position.
 In the tensor product with the state the  crystal operator 
$\tau_{\alpha}$ ($\alpha = H, V$) will be considered in the second 
position. A codon, e.g. XZN, will be considered subject to a 
translational error, 
e.g. to be read as XZN', if the crystal operator modelising the relevant 
translational error will connect, in the sense above explained, the 
state  $\psi(XZN)$ with the state  $\psi(XZN')$  where
$\psi(XZN) $ is the state in the
 irreps. of $U_{q \to 0}(sl_{H}(2) \oplus sl_{V}(2)) $, see  Table 
\ref{tablerep}, specifying the codon XZN. 
  
\subsection{ Substitution of 3rd  nucleotide}

To study the  transitions in 3rd position in the codons XZC and XZG, 
where X and Z are any nucleotide, we consider the action of the operator 
given by eq.(\ref{eq:t1})  with $a = 0$ on  the corresponding states:  
(in the following the equations have to be read by the western 
rule from left to right) 
\bea
\psi(XZC)  \circ (\tau_{H,-1}^{1} \otimes \tau^{0}_{V,0})  \; & \Lra  
& \; 
\psi(XZU)   \label{eq:3t1}  \\ 
\psi(XZG)  \circ ( \tau_{H,-1}^{1} \otimes \tau^{0}_{V,0}) \; & \Lra  
& \; 
\psi(XZA)  \label{eq:3t2}
\eea
 We impose   the codons $XZC$ (resp.  $XZG$)  and $XZU$ (resp.  $XZA$) 
 to be synonymous, if the states are connected by the
  $ \tau_{H,-1}^{1}  \otimes \tau^{0}_{V,0}$ 
 according to the ($q \to 0$)-Wigner-Eckart theorem. We get  the 
splitting of the 64 codons  in 32 doublets of the form XZR and XZY.
Remark that the final  pattern  is unchanged if in eq.(\ref{eq:3t1})
we replace $\tau^{0}_{V,0}$ by $\tau^{1}_{V,0}$. 
To study the  tranversions in 3rd position in the codons XZC and XZU 
we consider the  action of  the crystal operators given by 
eq.(\ref{eq:tr1}) and  eq.(\ref{eq:tr2})   on  the corresponding 
states:  
\bea
 \psi(XZC)  \circ (\tau_{H,0}^{b}  \otimes \tau_{V,-1}^{1})  \; & 
\Lra  & 
\; \psi(XZG) \label{eq:tva}  \\ 
\psi(XZU)  \circ (\tau_{H,0}^{b - 1}  \otimes \tau_{V,-1}^{1})  \; & 
\Lra  & 
\; \psi(XZA)  \label{eq:tvb} \\ 
 \psi(XZC)  \circ (\tau_{H,-1}^{b}  \otimes \tau_{V,-1}^{1} ) \; & 
\Lra  & 
\; \psi(XZA) \label{eq:tvc}
\eea  
where in eqs.(\ref{eq:tva}),(\ref{eq:tvb}),(\ref{eq:tvc})
 $b = 2$   if  the dinucleotide XZ, i.e. the state formed by the 
first two nucleotides in the initial codon, belongs to an irrep.
with $J_{V} = 0$ or is a state with lowest weight for
$sl_{V}(2)$  or for $sl_{H}(2)$, if $J_{H} \neq 0$, \cite{FSS1}, see 
Table \ref{tabledin} (i.e.
 when the first two nucleotides are: CA, GA, CG, UG, UA, UU, AU, AA, 
 GG, AG)  and $b = 1$   otherwise.
 We have previously given arguments to motivate the introduction
 of different tensor operators to describe the  misreading of the same
 nucleotide in different codons, it is just a simple way of mathematically
 mimicking the dependence of the translational errors or misreading
 from the neighbouring nucleotides. For the transitions, which imply
  errors in the translation betwen members of the same chemical family, the tensor 
 operators depend only from the position in the codons of the misread
 nucleotide.  Let us comment more on the mathematical meaning of our
 assumptions. Due to the peculiar properties of the crystal basis 
  the tensor operators  can be assumed to consist of a part, let us
  say $ \eta $, with definite transformations 
 properties with respect to the generators of  $U_{q \to 0}(sl(2) \oplus 
sl(2))$  acting {\bf on the first two nucleotides}  and of a part, let us
say $ \tau$, with definite transformations 
 properties with respect to the generators of  $U_{q \to 0}(sl(2) \oplus 
 sl(2))$ acting on the whole codon. Under the action of $\eta$ the state of initial  
  codon changes into a "virtual" state transformed by $\tau$ into
  a final state. If the labels of the final state correspond to the labels of 
  the state describing the codon, the transversion is induced, otherwise it
  is not allowed, i.e. there is no misreading. This kind of reasoning is 
  applied in the case of substitution of two nucleotides, see Subsection 3.4. The choice of a different
  rank of  $\tau_{H,0}^{b}$ in eqs., (\ref{eq:tva}), (\ref{eq:tvb}), 
  (\ref{eq:tvc}) is a simple way to take into account this complex mechanism.
  One may reformulate the above conditions as it follows: 
 in eqs.(\ref{eq:tva}),(\ref{eq:tvb})     $b = 2$ 
  if the codon and the one obtained by transversion 
belong to the same irrep.  or if the initial codon belongs to an 
irrep. with $J_{H} = \Half$ and $b = 1$  otherwise;  in 
eq.(\ref{eq:tvc}) $b = 2$, if the codons XZC and XZU  belong to the same 
irrep.and $b = 1$  otherwise. These conditions are  simpler, but the dependence of the 
rank of the operators also from the irrep. of the final codon may sound unsatisfactory.
 It turns out that: 
\begin{itemize} 
\item   eq.(\ref{eq:tva}) forbids  transversions UUC $ \ra $ UUG, 
AUC $ \ra $ AUG, AAC $ \ra $ AAG, UAC $ \ra $ 
UAG,  GAC $Ê\ra $ GAG,  and CAC $Ê\ra $ CAG;
\item eq.(\ref{eq:tvb}) forbids transversions UUU $ \ra $ UUA, 
AUU $ \ra $ AUA, AAU $ \ra $ AAA, UAU $ \ra $ 
UAA, AGU $ \ra $ AGA,  GAU $Ê\ra $ GAA, UGU $Ê\ra $ UGA and CAU $Ê\ra $ CAA
\item eq.(\ref{eq:tvc})  forbids transversions UUC $ \ra $ UUA, 
AUC $ \ra $ AUA, AAC $ \ra $ AAA, UAC $ \ra $ 
UAA, AGC $ \ra $ AGA,  GAC $Ê\ra $ GAA, UGC $Ê\ra $ UGA and CAC $Ê\ra $ 
CAA 
\end{itemize}
Therefore we obtain   the  merging of 16 doublets in 8 quartets,
 the  quartets being the codons whose the first two 
 nucleotides are: CC, CU, CG, UC, GG, GC, GU, and AC.
 Let us note that the transitions AGC $ \ra $ AGG and UGC $ \ra $ UGG  are 
 allowed; a way of insuring protection without decreasing the number of
 amino acids encoded is to  make an appropriate choice for the  codons AGG 
and UGG 
 in the encoding process; indded in VMC the first is a  stop codon while 
 the second encodes for a a very rare amino acid Trp.
 At this stage the assignment of the codons, differing for the 3rd 
   nucleotide, to different multiplets is decided. The next steps 
   can produce the joining of doublets and quartets in  quartets or 
sextets or in, a priori, octets.  

Let us study what it is obtained 
if we change the mathematical modelisation of  the  direct transversions
( C $\ra$ G, U $\ra $ A) in 3rd position using the following   operator 
\bea
\psi(XZC) \circ (\tau_{H,O}^{\alpha} \otimes \tau_{V,-1}^{1}) \; & \Lra  &\;
\psi(XZG)   \label{eq:2tv3a} \\ 
\psi(XZU) \circ (\tau_{H,0}^{\beta}  \otimes \tau_{V,-1}^{1}) \; & \Lra  &\;
\psi(XZA)   \label{eq:2tv3b}
\eea
where:

\begin{itemize}

\item $\alpha =  2$  if the  state $\phi(XZ)$ of the dinucleotide XZ     
 is a lowest weight state for $sl_{H}(2)$ in an irrep. with $ J_{H} \neq 0 $ 
(i.e. from Table \ref{tabledin}: XZ = UU, AU, AA) 
 and  $\alpha = 1$   otherwise. 
\item   $\beta =  0$,   if the   dinucleotide
 is a state unmodified by the
action of a vector operator $\eta_{V,O}^{1} $ acting on it,  in the sense
that the labels of the state
\be
  \phi'(XZ) = \phi(XZ) \circ \eta_{V,O}^{1}
  \ee
  are the same than the state $\phi(XZ)$, and $ \beta = 1$ otherwise, i.e.
  from Table \ref{tabledin}: XZ = CU, GU, CC, UC, UU, GC, AC, AU.
 \end{itemize}
   
 In the spirit of the  hierarchical strength  of misreading errors, 
 a quartet will be formed surely if both the {\bf codons XZR} are transformed in 
 {\bf XZY}. It turns out the  merging of 16 doublets in 8 quartets,
 the  quartets being the codons whose the first two 
 nucleotides are: CC, CU, CG, UC, GG, GC, GU, and AC.  It turns out also 
 that: 
 \begin{itemize}
 \item in AGC, URC, CAC and GAC    the nucleotide C in 3rd position  
  can be transformed in  G while in AGU and UGU  the nucleotide U in 3rd
  position  cannot be transformed in A
  \item in UUU and AUU the U in the end position can be transformed in A, while 
  in UUC and AUC the C cannot be transformed in U.
  \end{itemize}
Let us analyse more in detail the function and physical-chemical properties 
of the  doublets in which only one state is subjetc to  misreading .
We remark that UAY and AGY encode in VMC the stop codons. Moreover the 
the physical-chemical properties of His (encoded vy CAY), Asp (GAY), Cys 
(UGY) and Asn (AAY) are, respectively, close to the properties pf Gln 
(CAR), Glu (GAR), Trp (UGR) and Lys (AAR). (\footnote  {For an explanation of
this affinity, which is indeed observed, in the framework of the crystal 
basis model see \cite{FSS2}}). Moreover in SUC there is a breaking 
of the doublet AUY, AUA merging with the doublet AUY in a triplet encoding for Ile.
It is tempting to draw the conclusion that, when the push to form a larger 
multiplet    acts only on some codons,  the nature seems to choose to have
a larger variety of a.a.  choosing the codons subject to misreading or as stop
 or  to encode affine a.a..
     As a final remark,  modelising the transversions simply by
     the vector operator $\tau_{H,O}^{1} \otimes \tau_{V,-1}^{1}$  in
     eqs.(\ref{eq:2tv3a})-(\ref{eq:2tv3b}) we  obtain the clear  merging 
     of  eight doublets in four
     quartets (CCN, CGN, GCN, GGN), which are indeed the "strongest" 
     quartets involving a triple hydrogen bond.

  \subsection { Substitutions of 1st  nucleotide}

We study first the transitions using the crystal vector operators 
introduced in eq.(\ref{eq:t1}) with $b = 1$  acting on the 
first nucleotide. So we study the transition  
\bea
\psi(CXN)  \circ (\tau_{H,-1}^{1} \otimes \tau^{1}_{V,0})  \; & \Lra  
& \; 
\psi(UXN)  \nonumber \\
\psi(GXN)  \circ (\tau_{H,-1}^{1} \otimes \tau^{1}_{V,0})  \; & \Lra  
& \; 
\psi(AXN) \label{eq:to1}
\eea
  One computes that only the following transitions are allowed:
  \begin{enumerate}
  \item in the quartets CUN and CCN,  for the states with N = U, A 
\item in the quartets  CGN, GGN, GCN and GUN  and in the doublets CAY 
and GAY for the states with N = U , Y = U 
\end{enumerate}    
According to the strategy of protection of the "weakest" codon above 
outlined, a fusion of a doublet with a quartet in a sextet or with another 
doublet  in a quartet (resp. of two quartets into an octet) happens if at 
least the transition of the codon with C or A in 3rd position (resp. of 
the 
codons with C and A in 3rd position) is allowed. In the light of the above 
criterion only the merging of the doublet UUR and the quartet CUN in a 
sextet is 
expected  and, indeed, we obtain the sextet encoding Leu.  Then let us 
analyse the tranversions in first position
\bea
\psi(CXZ)  \circ (\tau_{H,0}^{1} \otimes \tau_{V,-1}^{1})  \; & \Lra  
& \; 
\psi(GXZ) \label{eq:1tv1}   \\ 
\psi(UXZ)  \circ (\tau_{H,0}^{2} \otimes \tau_{V,-1}^{1})  \; & \Lra  
& \; 
\psi(AXZ) \label{eq:1tv2}  \\    
\psi(CXZ) \circ (\tau_{H,-1}^{c} \otimes \tau_{V,-1}^{1} ) \; & \Lra  
& \; 
\psi(AXZ)   \label{eq:1tv3}
\eea
where $c = 1$ if the codons CXZ and UXZ  belong to the same irrep.
  and $c = 2$ otherwise.   
 It turns out:
\begin{itemize} 
\item eq.(\ref{eq:1tv1}) allows only the transversions CCG $ \ra $ 
GCG, CCA $ \ra $ GCA, CGA $ \ra $ GGA, CAG $ \ra $ GAG and CGG $ \ra $ GGG  
\item  eq.(\ref{eq:1tv2}) allows only the transversions UCG $ \ra $ 
ACG and UGG $ \ra $ AGG  
\item  eq.(\ref{eq:1tv3}) allows only the tranversions CCA $ \ra $ 
ACA,CGA $ \ra $ AGA,  CUG $Ê\ra $ AUG  and  CAG $Ê\ra $ AAG.
\end{itemize}
 As a consequence the doublet AGR  merge into the quartet CGN forming 
  another sextet encoding for Arg.
 Remark that the established pattern remains unchanged if the rank of  $\tau_{H}$ in
 both eqs.(\ref{eq:1tv1})-(\ref{eq:1tv2})  is fixed 1 or 2. 
  So at this stage the multiplet structure of the 64 codons is:
   2 sextet, 6 quartets and 14 doublets, 2 of which are splitted in
   singlets. We get almost the structure  of the VMC or of SUC, the 
Ser sextet being missed. 
 
 An alternative mathematical scheme to modelise the transition is:  
 \bea
\psi(CXN)  \circ (\tau_{H,-1}^{1} \otimes \tau^{0}_{V,0})  \; & \Lra  
& \; 
\psi(UXN)  \nonumber \\
\psi(GXN)  \circ (\tau_{H,-1}^{1} \otimes \tau^{0}_{V,0})  \; & \Lra  
& \; 
\psi(AXN) \label{eq:2to1}
\eea
  One computes that the following transitions happen:
  \begin{enumerate}
  \item in the quartets CCN, CGN, CUN, GGN, GCN and GUN  for the states with N = U, A 
\item in the doublets CAR, CAY, GAR and GAY for the states with R = A, Y = U  
\end{enumerate} 
   
 According to the strategy above outlined, we expect:
 \begin{enumerate}
 \item  the fusion of the doublets CAR and UAR  and GAR and AAR respectively in two 
 quartets 
 \item   the fusion of the doublets UUR (resp. UGR, AUR, AGR) and the quartets 
 CUN (resp. CGN, GUN, GGN) in sextets.
 \end{enumerate}  
  A way to satisfy the stability condition without
 decreasing the number of a.a. synthetized is to make an 
 appropriate choice  for the stop codons, as already remarked. In fact the 
 decreasing of an encoded a.a. is avoided choosing UAR as stop codons. Moreover  
   the doublet UUR merges with the quartet CUN in the sextet
 encoding Leu, while in VMC UGR encodes a rare a.a. Trp  and in SUC the doublet is
 splitted in two singlets, UGG encoding  Trp and UGA, state subject to 
 mutation, encoding Ter. The fusion of AGR with GGN  does not happen, but in VMC this 
 doublet encodes for stop codons and in SUC merges in another sextet, as we shall 
  see below, while the quartet GAR and AAR is not found. However it is worth to note 
  that some physical-chemical properties of the two encoded a.a. (Glu and Lys)
   are very close, see \cite{FSS2}, and that the two codons are formed
   only by purine, with prevailing  nucleotide A. May be  also that  the
   requirement of the merging of two doublets into a quartet  when only the 
   codon with a A nucleotide in the final position is subject to error, is
   a too strong condition. 
    So we have found further arguments   in favour of  UAY  and AGY  
    being stop codons.
    Then let us analyse the tranversion 
 $ C \ra A$ which can be read as the result of $ C \ra G \ra A$ due to
 the tensor operator 
 \be 
 \tau_{H,-1}^{1} \oplus \tau_{V,-1}^{1}  
 \ee 
  It turns out that only the transversion  CXA $ \ra$ AXA is allowed.
  As a consequence we expect the merging of the doublet AGR with the quartet 
  CGN, of the doublet AUR with the quartet CUN and, eventually, of the two doublets AAR 
  and CAR.  
  Only the first sextet is observed, but the doublet AUR encodes
  the starting codon in VMC and is split out in SUC.

 \subsection { Substitution  of  central nucleotide}
   
   The translation errors in the 2nd nucleotide occur very rarely, so 
  we consider it as  weak intensity effect,  assuming that 
  it cannot modify the already established pattern in doublets and 
  quartets, but only to possibly cause the merging of whole 
multiplets.
   We modelise the transitions as   
\bea
\psi(XCN)  \circ (\tau_{H,-1}^{1} \otimes \tau^{2}_{V,0})  \; & \Lra  
& \; 
\psi(XUN)  \nonumber  \\
\psi(XGN)  \circ (\tau_{H,-1}^{1} \otimes \tau^{2}_{V,0})  \; & \Lra  
& \; 
\psi(XAN) \label{eq:to2}
\eea
 From the results of previous subsections we know that the codons 
with  C or G in first  position and  C or U  in the central position are 
organised in  quartets, therefore only an octet is the possible larger 
multiplet. According to the general  strategy followed, the fusion of two 
quartets is possible if at least the following transitions VCK $ \ra $ VUK 
(V = C, G; K = C, A) are allowed. For the codons with U or A in first 
position and C in second position  the fusion  of a quartet WCK (W = U, A) 
and a doublet WUR (resp. WUY)  in a sextet is possible if   at least the 
transition  WCA $ \ra $ WUA (resp. WCC $ \ra $ WUC) is possible. The 
fusion in 
sextet of a  quartet VGN (V = C, G) with a doublet VAR (resp. VAY) is 
possible if 
 at least the transition  VGA $ \ra $ VAA (resp. VGC $ \ra $ VAC).
Finally the fusion of two doublets WGR  and WAR (W = U, A) (resp. WGY 
and WAY) should take place if at least the transition  WGA $ \ra $ WAA 
(resp. WGC $ \ra $ WAC) is allowed.  It turns out that all the above 
listed transitions are forbidden. Indeed only the transitions  
   CCC $ \ra $  CUC and GCC $ \ra $ GUC are allowed.  
     
      Let us analyse the tranversions in  second position
    \bea
\psi(XCZ)  \circ (\tau_{H,0}^{1} \otimes \tau_{V,-1}^{2})  \; & \Lra  
& \; 
\psi(XGZ)    \label{eq:2tv1} \\
\psi(XUZ)  \circ (\tau_{H,0}^{2} \otimes \tau_{V,-1}^{2})  \; & \Lra  
& \; 
\psi(XAZ)    \label{eq:2tv2} \\ 
\psi(XCZ) \circ (\tau_{H,-1}^{c} \otimes \tau_{V,-1}^{2} ) \; & \Lra  
& \; 
\psi(XAZ)   \label{eq:2tv3}
 \eea
where $c = 1$ if the codons XCZ and XUZ belong to the same   
irrep.  and $c = 2$ otherwise.
 It turns out:
\begin{itemize} 
\item eq.(\ref{eq:2tv1}) allows  the transversions MCC $ \ra $ MGC 
($ M \neq C$) 
\item  eq.(\ref{eq:2tv2}) does not allow any   transversion 
\item  eq.(\ref{eq:2tv3}) allows only the tranversions CCC $ \ra $ 
CAC, UCU $ \ra $ UAU  and  ACU $Ê\ra $ AAU.
\end{itemize} 
 It turns out that one should expect the fusion in a sextet of the 
quartet UCN and of the doublet UGY as the transition UCC $ \ra$ UGC is 
allowed.This sextet does not appear in the genetic code, but as we shall 
see in the following subsection indeed the quartet UCN merges with the 
 doublet AGR. One should also expect the fusion in a sextet of the quartet 
CCN and the doublet CAY, which indeed does not happen. 
Both these results suggest that the misreading of the central nucleotide
is a very weak effect, if not enhanced by the simultaneous misreading of 
the first nucleotide, see the following Subsection.
Remark that in eq.(\ref{eq:2tv1})  we might write $\tau_{H,0}^{2}$ which 
leaves the final result unchanged (with this choice also the transition
CCC $ \ra $ CGC is allowed).

 \subsection{ Substitution of two nucleotides}
  
  The reading errors in a couple of  nucleotides is an event
  occurring less frequently than the translation errors of one
  nucleotide in last or initial position, therefore we  generally
  expect a weaker effect than the previously considered one 
nucleotide 
  change. Consequently  we assume that they cannot
  modify the already established pattern in doublets and quartets.
  So we consider only the possible action on the two initial 
  nucleotides. The transition  and tranversion   of the first
   (second)  nucleotide  is modelised by the same operator used  
    for the translation or transversion on the first nucleotide,
   see eqs.(\ref{eq:to1}),(\ref{eq:1tv1}),(\ref{eq:1tv2}),(\ref{eq:1tv3})
   (see eqs.(\ref{eq:to2}),(\ref{eq:2tv1}),(\ref{eq:2tv2}), 
   (\ref{eq:2tv3})). 
    In the following we denote with a lower label the position of 
   the nucleotide where the operator acts.
   The action of the  two-nucleotides operators  has to be computed in
    the following way: as first step one has to compute the action of
 the operator labeled by I giving rise to a "virtual" state with the
    labels assigned by the action of the relevant operator on the 
    initial state of the codon, then one considers the action of the 
    operator labeled by II on the "virtual" state and gets the  labels
    of the final state. If these labels are the ones denoting in Table \ref{tablerep} 
    the state corresponding to the codon, the transition is allowed. 
    For example to analyse the transition CCN in UUN one should 
compute
    \bea
    & \psi(CCN) \circ (\tau_{H,-1}^{1} \otimes \tau_{V,0}^{1})_{I} \, \ra 
\,  \psi((UCN)_{v})
    \nonumber \\
     & \psi((UCN)_{v}) \circ (\tau_{H,-1}^{1} \otimes \tau_{V,0}^{1})_{II} 
\, \ra  \, \psi(UCN) 
     \eea 
     where the labels of the state  $(UCN)_{v}$ are computed by
     \be
     \psi((UCN)_{v}) = \psi(CCN) \otimes (\tau_{H,-1}^{1} \otimes 
\tau_{V,0}^{1})_{I} 
    \ee
 It follows that one can get an allowed transition and/or transversion  to a final state, 
 even if the action of the operator labelled by I does not induce  it. The 
same kind of computation has to be performed in the cases of transition + 
tranversion or viceversa or double tranversion.  Let us analyse:
    \begin{itemize}
    \item the double transitions 
   
   ($CCN \ra UUN$, $GGN \ra AAN$, $CGN \ra UAN$, $GCN \ra AUN $)   
   \be 
     (\tau_{H,-1}^{1} \otimes \tau_{V,0}^{1})_{I} \; \oplus  \; 
     (\tau_{H,-1}^{1}  \otimes \tau_{V,0}^{2})_{II}   
    \label{eq:TT}
    \ee
 Only the transitions CCU $ \ra$ UUU  and GCU $ \ra$ AUU  are allowed.
 Note that obtained pattern is unmodified if we modelise the double 
 transition by
 \be 
     (\tau_{H,-1}^{1} \otimes \tau_{V,0}^{0})_{I} \; \oplus  \; 
     (\tau_{H,-1}^{2}  \otimes \tau_{V,0}^{0})_{II}   
  \ee
 
 \item the transition + tranversion:
    
    ($CCN \ra UGN$,  $CUN \ra UAN$, $GCN \ra AGN$,  $GUN \ra AAN$)  
    \be 
     (\tau_{H,-1}^{1} \otimes \tau_{V,0}^{1})_{I} \; \oplus  \; 
     (\tau_{H,0}^{b} \otimes \tau_{V,-1}^{2})_{II}   
    \label{eq:TTV}
    \ee
   where here and in the following $b = 1$ ($b = 2$) for transversion C $\ra$ G  (U $\ra$ A).
   Only the transition CUC $ \ra $ UAC is allowed.
    
   ($CCN \ra UAN$, $GCN \ra AAN$)  
    \be 
     (\tau_{H,-1}^{1} \otimes \tau_{V,0}^{1})_{I} \; \oplus  \; 
     (\tau_{H,-1}^{2} \otimes \tau_{V,-1}^{2})_{II}   
    \label{eq:TTVD}
    \ee
   Only the transitions CCU $ \ra $ UAC and GCU $ \ra $ AAU are allowed.  
    
    \item transversion + transition:
    
    ($CCN \ra GUN$, $CGN \ra GAN$, $UCN \ra AUN$,  $UGN \ra AAN$)
     \be 
     (\tau_{H,0}^{b} \otimes \tau_{V,-1}^{1})_{I} \; \oplus  \; 
     (\tau_{H,-1}^{1} \otimes \tau_{V,0}^{2})_{II}   
    \label{eq:TVT}
    \ee
     Only the transversion-transitions CCY $ \ra $ GUY are allowed. 
    
    ($CCN \ra AUN$, $CGN \ra AAN$)  
     \be 
     (\tau_{H,-1}^{1} \otimes \tau_{V,-1}^{1})_{I} \; \oplus  \; 
     (\tau_{H,-1}^{1} \otimes \tau_{V,0}^{2})_{II}   
    \label{eq:TVTD}
    \ee
    Only the transversion-transition CCU $ \ra $ AUU is allowed. 
    
    \item double transversion:
      
    ($CCN \ra GGN$, $CUN \ra GAN$, $UUN \ra AAN$, $UCN \ra AGN$)
     \be 
     (\tau_{H, O}^{b} \oplus \tau_{V, -1}^{1}) _{I} \; \oplus  \; 
     (\tau_{H, 0}^{b} \oplus \tau_{V, -1}^{2})_{II}  
   \label{eq:TVTV}
   \ee
    Only the transversions UCC $ \ra $ AGC and CCC $ \ra $ GGC are allowed.
    
    ($CCN \ra AGN$, $CUN \ra AAN$)
   \be 
     (\tau_{H, -1}^{1} \oplus \tau_{V, -1}^{1}) _{I} \; \oplus  \; 
     (\tau_{H, 0}^{b} \oplus \tau_{V, -1}^{2})_{II}  
   \label{eq:TVTVD1}
   \ee 
   No transversion is induced.
   
   ($CCN \ra GAN$, $UCN \ra AAN$) 
     \be 
     (\tau_{H, O}^{b} \oplus \tau_{V, -1}^{1}) _{I} \; \oplus  \; 
     (\tau_{H, -1}^{2} \oplus \tau_{V, -1}^{2})_{II}  
   \label{eq:TVTVD2}
   \ee 
    Only the transversion CCG $ \ra $ GAG  is allowed.
    
   ($CCN \ra AAN$) 
 \be 
     (\tau_{H, -1}^{1} \oplus \tau_{V, -1}^{1}) _{I} \; \oplus  \; 
     (\tau_{H, -1}^{2} \oplus \tau_{V, -1}^{2})_{II}  
   \label{eq:TVTVDD}
   \ee 
    Only the transversion CCG $ \ra $ AAG  is allowed.
    \end{itemize}
    It turns out, using also the results  and the discussions of the 
previous subsections,  that the action of the above operators does not 
modify the 
established pattern  except the one given by eq.(\ref{eq:TVTV}) which 
induces the mutation 
 $ UCA \ra AGA$, so  urging  the doublet AGR to merge with the
  quartet UCN  giving rise to the third sextet encoding for Ser.

\section{Conclusions}
  
  Before discussing what we have obtained, let us summarize what we
  have done. The starting point is the  observed pattern in 
  multiplets  of the genetic code. From  its invariance in time
  and from its, almost, universal character we infer that such a
  pattern has to ensure an efficient and stable translation in
  the building of polypeptides chains, i.e. it is  error proof
  against the most frequent reading errors. To  give a quantitative 
and precise meaning to this statement we need to build a mathematical 
  model both for the genetic code and  for the misreading mechanisms.
  In the crystal basis model each codon is represented as a state 
  $\psi_{(J_{H}, J_{V}; J_{H,3}, J_{V,3})}$ in 
  the module space of $U_{q \to 0}(sl(2) \oplus sl(2))$. The 64 states
  are separated in nine different invariant subspaces labelled  by a couple of
  half-integer $J_{H}, J_{V}$. The mechanisms implying translation
  errors are modelised by suitable tensor operator, with definite
  transformation properties under $U_{q \to 0}(sl(2) \oplus sl(2))$,
  which may or may not relate two states of such states.
  If the states are connected, we infer that they can be mistaken in
  the translation process, and therefore, in order to ensure in case
  of misreading the synthesis of the same a.a., the corresponding
  codons have to be synonimous. By studying the action of the
  operators, we obtain  the splitting of the 64
  states in a set of multiplets representing almost faithfully the   
  degeneracy of the genetic code.
  The simple proposed mathematical modelisation    is able, in an 
  amazing way, to account almost for the existence af only 20 a.a. 
  and, almost, for the structure of the VMC and SUC. Why 
  the nature uses the 20 particular 
  a.a., enumerated in the beginning of this paper, in the practically
  unlimited variety of these molecules is still to be understood  
  and, of course, is far beyond the aim of this work.
   The structure of the mathematical operators used to model 
transitions and tranversions is  simple but arbitrary.
  Therefore it is worth to discuss in a more quantitative way the
  extent of the obtained results.
  Let us use as starting point the pattern of the 64 codons grouped 
  in 32 doublets, even if this result is less obvious that one can
  naively think of. Indeed from Table \ref{tablerep} one realizes 
  that 8 of the 16 doublets of the form XZY (resp. XZR) belong to 
different irreps. Therefore the vector operator 
  given in eqs.(\ref{eq:3t1}), (\ref{eq:3t2})  operates
  in half the case as a generator and in half case as an intertwining
  operator. The transversion operator induces the merging of 16 
doublets in 8 quartets in full agreement with the observed pattern of the 
genetic code. The formation of only 8 quartets, with the correct content in 
  the first two nucleotides, induced by the action of operator 
  eqs.(\ref{eq:tva}), (\ref{eq:tvb}), (\ref{eq:tvc})   is a good result,
   especially considering that
  the number of different choices of 8 quartets in 16 doublets is
  12870. Once formed the quartets the operator eq.(\ref{eq:to1})
  induces the formation of 2 sextets which are the correct ones 
between the 420 possibilities. Finally the operator eq.(\ref{eq:TVTV}) 
induces the formation of the correct 3rd sextet betwwen 24 possibilities.
  In conclusion it is  extremeky surprising that such
  an arbitrary choice explains why and in which pattern of multiplets (with a 
probability to find the correct pattern of about $ 7,7 \cdot 10^{-9}$) the 
remaining 60 codons encode   only 20 amino acids. 
 We have invenstigated the dependence of the pattern obtained from the
  structure of the tensor operators used to modelise the misreading 
  process. Differences do appear  in the different   modelisations studied, but
    most of the pattern of
   the genetic code is obtained, showing that there is a  bulk of its organisation 
 little sensitive to the details of the operators modelising the 
 misreading process. This feature 
 appears also in other modelisation not discussed in the  paper, e.g.
 modelising the transition and transversion as a two steps process:
 deletion of a nucleotide and subsequently creation of a different one.
 A very few differences, depending also from the choosen scheme, exist between 
 the theoretical pattern of organisation in multiplets and the observed one.
 In particular some  minor changes 
 in the eukaryotic code do not find an explanation in the model, even if 
 for some of them the model give hints in the correct direction.  Furthere refinements or,
  more probably, the 
presence of  some other  mechanism whose action is not modelisable by crystal 
tensor operators may account  for these changes and for the not 
appearance of an expected  4th sextet in the second mathematical scheme. 
However it should remarked
  that this sextet is formed by CCN and CAY  where CC (resp. CA) is the highest
   weight (resp. the lowest  weight) in the dinucleotide set.
 
  In our model the strategy followed by the genetic code
  seems  to  be adressed to keep the most variety of encoded amino
  acids consistently with a reasonable level of protection of the 
  codons against the most common translation errors.
    A fundamental problem, not all faced in this paper, is the reason
    for the observed correspondence between multiplets and amino acids;
    in other words once obtained the organisation in different multiplets
    of the genetic code,  there is a  mechanism imposing  which particular amino 
    acids have to be encoded by sextets, quartets and so on or it is just
    a random event ?
      Stereochemical hypothesis  \cite{Wo}, \cite{S} suggests that the physical-chemical 
     properties of the amino acids play a crucial role to determine the
     correspondennce between multiplets and amino acids. 
 A clear shortcoming of the model is the fact 
  that the analysis of the formation of the different multiplets is 
peformed in a "static" manner while it is believed, although no unambigous 
model does exist, that an evolution of the genetic code and of the 
  corresponding encoded amino acids has happened. It is indeed in the 
  evolution that rules and properties of the systems which are 
  necessary to the existence of living organisms  are fixed and 
  selected.   The selection, which dominates biology, 
has not at all taken into account in the present oversimplified model.
  However  hopefully this kind of  reasoning can be applied to models 
   describing the evolution process.
      One may argue that, in a more refined model, different 
   operators should be used to modelise different mutagenic effects,
   whose role and intensity depend on the in time changing 
environment. We point out also that the one may conjecture to modelise
spontaneous and induced mutations  of  the genetic code by suitable tensor 
operators, a first analysis of this type has been  given in \cite{FSS3}. 
In conclusion the model presented in 
   this paper states that the genetic code is what it is because it is
  "optimized", at least for the environment in which it was formed, 
and not for a freezing random event. Of course the word optimization should 
be taken in a loose sense as we have not quantitatively described the
  gain of the different choice. We believe that in this context methods of game theory
  can be appropriately used to a better description.
 
  \bigskip
  
   \textbf{Acknowledgments}:  I thank M. Di Giulio for discussions and very
 helpful  suggestions.

\clearpage

\begin{table}[htbp]
\caption{The vertebral mitochondrial code. The upper label denotes 
different irreducible
representations. In bold character the amino acids which are encoded 
dfferently  in the eukariotic or standard code: UGA, AUA, AGY encoding
 respectively for Ter, Ile and Arg.  }
\label{tablerep}
\footnotesize
\begin{center}
\begin{tabular}{|cc|cc|rr|cc|cc|rr|}
\hline
codon & a.a. & $J_{H}$ & $J_{V}$ & $J_{3,H}$ & $J_{3,V}$& codon & 
a.a. & 
$J_{H}$ & $J_{V}$ & $J_{3,H}$ & $J_{3,V}$ \\
\hline
\Big. CCC & Pro & $\Half$ & $\Half$ & $\Half$ & $\Half$ & UCC & Ser & 
$\Half$ & 
$\Half$ & $\half$ & $\Half$ \\
\Big. CCU & Pro & $(\half$ & $\Half)^1$ & $\half$ & $\Half$ & UCU & 
Ser & 
$(\half$ & $\Half)^1$ & $-\half$ & $\Half$ \\
\Big. CCG & Pro & $(\Half$ & $\half)^1$ & $\Half$ & $\half$ & UCG & 
Ser & 
$(\Half$ & $\half)^1$ & $\half$ & $\half$ \\
\Big. CCA & Pro & $(\half$ & $\half)^1$ & $\half$ & $\half$ & UCA & 
Ser & 
$(\half$ & $\half)^1$ & $-\half$ & $\half$ \\[1mm]
\hline
\Big. CUC & Leu & $(\half$ & $\Half)^2$ & $\half$ & $\Half$ & UUC & 
Phe & 
$\Half$ 
& $\Half$ & $-\half$ & $\Half$ \\
\Big. CUU & Leu & $(\half$ & $\Half)^2$ & $-\half$ & $\Half$ & UUU & 
Phe & 
$\Half$ 
& $\Half$ & $-\Half$ & $\Half$ \\
\Big. CUG & Leu & $(\half$ & $\half)^3$ & $\half$ & $\half$ & UUG & 
Leu & 
$(\Half$ & $\half)^1$ & $-\half$ & $\half$ \\
\Big. CUA & Leu & $(\half$ & $\half)^3$ & $-\half$ & $\half$ & UUA & 
Leu & 
$(\Half$ & $\half)^1$ & $-\Half$ & $\half$ \\[1mm]
\hline
\Big. CGC & Arg & $(\Half$ & $\half)^2$ & $\Half$ & $\half$ & UGC & 
Cys & 
$(\Half$ & $\half)^2$ & $\half$ & $\half$ \\
\Big. CGU & Arg & $(\half$ & $\half)^2$ & $\half$ & $\half$ & UGU & 
Cys & 
$(\half$ & $\half)^2$ & $-\half$ & $\half$ \\
\Big. CGG & Arg & $(\Half$ & $\half)^2$ & $\Half$ & $-\half$ & UGG & 
Trp & 
$(\Half$ & $\half)^2$ & $\half$ & $-\half$ \\
\Big. CGA & Arg & $(\half$ & $\half)^2$ & $\half$ & $-\half$ & UGA & 
{\bf Trp} & 
$(\half$ & $\half)^2$ & $-\half$ & $-\half$ \\[1mm]
\hline
\Big. CAC & His & $(\half$ & $\half)^4$ & $\half$ & $\half$ & UAC & 
Tyr & 
$(\Half$ & $\half)^2$ & $-\half$ & $\half$ \\
\Big. CAU & His & $(\half$ & $\half)^4$ & $-\half$ & $\half$ & UAU & 
Tyr & 
$(\Half$ & $\half)^2$ & $-\Half$ & $\half$ \\
\Big. CAG & Gln & $(\half$ & $\half)^4$ & $\half$ & $-\half$ & UAG & 
Ter & 
$(\Half$ & $\half)^2$ & $-\half$ & $-\half$ \\
\Big. CAA & Gln & $(\half$ & $\half)^4$ & $-\half$ & $-\half$ & UAA & 
Ter 
& 
$(\Half$ & $\half)^2$ & $-\Half$ & $-\half$ \\[1mm]
\hline
\Big. GCC & Ala & $\Half$ & $\Half$ & $\Half$ & $\half$ & ACC & Thr & 
$\Half$ & 
$\Half$ & $\half$ & $\half$ \\
\Big. GCU & Ala & $(\half$ & $\Half)^1$ & $\half$ & $\half$ & ACU & 
Thr & 
$(\half$ & $\Half)^1$ & $-\half$ & $\half$ \\
\Big. GCG & Ala & $(\Half$ & $\half)^1$ & $\Half$ & $-\half$ & ACG & 
Thr & 
$(\Half$ & $\half)^1$ & $\half$ & $-\half$ \\
\Big. GCA & Ala & $(\half$ & $\half)^1$ & $\half$ & $-\half$ & ACA & 
Thr & 
$(\half$ & $\half)^1$ & $-\half$ & $-\half$ \\[1mm]
\hline
\Big. GUC & Val & $(\half$ & $\Half)^2$ & $\half$ & $\half$ & AUC & 
Ile & 
$\Half$ 
& $\Half$ & $-\half$ & $\half$ \\
\Big. GUU & Val & $(\half$ & $\Half)^2$ & $-\half$ & $\half$ & AUU & 
Ile & 
$\Half$ 
& $\Half$ & $-\Half$ & $\half$ \\
\Big. GUG & Val & $(\half$ & $\half)^3$ & $\half$ & $-\half$ & AUG & 
Met & 
$(\Half$ & $\half)^1$ & $-\half$ & $-\half$ \\
\Big. GUA & Val & $(\half$ & $\half)^3$ & $-\half$ & $-\half$ & 
AUA & {\bf Met} & 
$(\Half$ & $\half)^1$ & $-\Half$ & $-\half$ \\[1mm]
\hline
\Big. GGC & Gly & $\Half$ & $\Half$ & $\Half$ & $-\half$ & AGC & Ser 
& 
$\Half$ & 
$\Half$ & $\half$ & $-\half$ \\
\Big. GGU & Gly & $(\half$ & $\Half)^1$ & $\half$ & $-\half$ & AGU & 
Ser & 
$(\half$ & $\Half)^1$ & $-\half$ & $-\half$ \\
\Big. GGG & Gly & $\Half$ & $\Half$ & $\Half$ & $-\Half$ & AGG & 
{\bf Ter} & $\Half$ & 
$\Half$ & $\half$ & $-\Half$ \\
\Big. GGA & Gly & $(\half$ & $\Half)^1$ & $\half$ & $-\Half$ & 
AGA & {\bf Ter} & 
$(\half$ & $\Half)^1$ & $-\half$ & $-\Half$ \\[1mm]
\hline
\Big. GAC & Asp & $(\half$ & $\Half)^2$ & $\half$ & $-\half$ & AAC & 
Asn & 
$\Half$ 
& $\Half$ & $-\half$ & $-\half$ \\
\Big. GAU & Asp & $(\half$ & $\Half)^2$ & $-\half$ & $-\half$ & AAU & 
Asn 
& $\Half$ 
& $\Half$ & $-\Half$ & $-\half$ \\
\Big. GAG & Glu & $(\half$ & $\Half)^2$ & $\half$ & $-\Half$ & AAG & 
Lys & 
$\Half$ 
& $\Half$ & $-\half$ & $-\Half$ \\
\Big. GAA & Glu & $(\half$ & $\Half)^2$ & $-\half$ & $-\Half$ & AAA & 
Lys 
& $\Half$ 
& $\Half$ & $-\Half$ & $-\Half$ \\[1mm]
\hline
\end{tabular}
\end{center}
\end{table}

\begin{table}[htbp]
\caption{ Irreducible representations of the dinucleotide states (dinucl.) }
\label{tabledin}
\footnotesize
\begin{center}
\begin{tabular}{|c|cc|rr|c|cc|rr|}
\hline
dinucl.  & $J_{H}$ & $J_{V}$ & $J_{3,H}$ & $J_{3,V}$& dinucl.  
  & $J_{H}$ & $J_{V}$ & $J_{3,H}$ & $J_{3,V}$ \\
 \hline 
  CC  & $1$ & $1$ & $1$ & $1$ & UC  & $1$ & $1$ & $0$ & $1$ \\
   CG  & $1$ & $0$ & $1$ & $0$ & UG  & $1$ & $0$ & $0$ & $0$ \\
    CU  & $0$ & $1$ & $0$ & $1$ & UU  & $1$ & $1$ & $-1$ & $1$ \\
     CA  & $0$ & $0$ & $0$ & $0$ & UA  & $1$ & $0$ & $-1$ & $0$ \\
     \hline \hline
      GC  & $1$ & $1$ & $1$ & $0$ & AC  & $1$ & $1$ & $0$ & $0$ \\
       GG  & $1$ & $1$ & $1$ & $-1$ & AG  & $1$ & $1$ & $0$ & $-1$ \\
        GU  & $0$ & $1$ & $0$ & $0$ & AU  & $1$ & $1$ & $-1$ & $0$ \\
         GA  & $0$ & $1$ & $0$ & $-1$ & AA  & $1$ & $1$ & $-1$ & $-1$ \\
\hline
 
\hline
\end{tabular}
\end{center}
\end{table}

\end{document}